\documentclass[a4paper,twoside]{article}

\usepackage{flushend}
\usepackage{subcaption}
\usepackage{graphicx}
\usepackage{xcolor}
\usepackage{url}
\usepackage{epsfig}
\usepackage{subcaption}
\usepackage{calc}
\usepackage{amssymb}
\usepackage{amstext}
\usepackage{amsmath}
\usepackage{amsthm}
\usepackage{multicol}
\usepackage{pslatex}
\usepackage{apalike}
\usepackage{graphicx}
\usepackage{SCITEPRESS}     

\begin{document}

\title{ZKlaims: Privacy-preserving Attribute-based Credentials using Non-interactive Zero-knowledge Techniques}

\author{\authorname{Martin Schanzenbach\sup{1}\orcidAuthor{0000-0001-6153-504X}, Thomas Kilian\sup{1}, Julian Sch\"utte\sup{1}\orcidAuthor{0000-0002-3007-6538} and Christian Banse\sup{1}\orcidAuthor{0000-0002-4874-0273}}
\affiliation{\sup{1}Fraunhofer AISEC, Parkring 4, Garching near Munich, Germany}
\email{\{schanzen, kilian, schuette, banse\}@aisec.fraunhofer.de}
}



\newcommand{\thething}{\mbox{ZKlaims}}

\abstract{In this paper we present \thething{}: a system that allows users to present attribute-based credentials in a privacy-preserving way.
We achieve a zero-knowledge property on the basis of Succinct Non-interactive Arguments of Knowledge (SNARKs).
\thething{} allow users to prove statements on credentials issued by trusted third parties.
The credential contents are never revealed to the verifier as part of the proving process.
Further, \thething{} can be presented non-interactively, mitigating the need for interactive proofs between the user and the verifier.
This allows \thething{} to be exchanged via fully decentralized services and storages such as traditional peer-to-peer networks based on distributed hash tables (DHTs) or even blockchains.
To show this, we include a performance evaluation of \thething{} and show how it can be integrated in decentralized identity provider services.}

\onecolumn \maketitle \normalsize \setcounter{footnote}{0} \vfill

\section{INTRODUCTION}

Recent events surrounding the (ab)use of personal identity information from
social networks once again rejuvenated the raison d'\^{e}tre of
privacy-preserving identity management~\cite{website:nytFacebookAnalytica}.
Classical attribute-based credential (ABC) systems such as X.509 certificates
raise privacy concerns as they reveal potentially sensitive attribute values
such as name, age, or social relationships to the credential verifier.
In order to alleviate this issue, privacy-preserving attribute-based credential
(PP-ABC) systems have been proposed in the past~\cite{paquin2011u,camenisch2002design}.
PP-ABCs rely on zero-knowledge protocols to prove statements over attributes,
rather than revealing the attribute values themselves.
Some systems require that prover and verifier must engage in an interactive
proving protocol which implies that prover and verifier must be online whenever
a credential verification is performed.
Other approaches remedy this issue by requiring that proofs of statements
are predefined by the credential issuer.
This inevitably requires the prover to interact with the issuer or verifier
whenever a genuinely new statement is required.
Both properties are particularly problematic in decentralized identity
management systems where centralized identity services are replaced by
peer-to-peer attribute provisioning.

Recent advancements in the area of non-interactive verifiable computation
schemes and the emergence of novel decentralized architectures such as
blockchains have paved the way for a new generation of decentralized,
privacy-preserving identity and access management~\cite{website:nameid,schanzen2016dpm,schanzenbach2018reclaim,decentid2018}.
A common component in such systems is a decentralized directory service used
to provision user attributes and credentials.
Such services are realized as a secure, shared storage medium such as a
blockchain or secure name system.
In order to fully leverage PP-ABCs, users must be able to present them
non-interactively over the shared medium.
Hence, to combine the privacy benefits of PP-ABCs with the privacy advantages
of decentralized identity management, non-interactive PP-ABCs are necessary.

\begin{figure*}[h]
  \centering
  \includegraphics[width=0.6\textwidth]{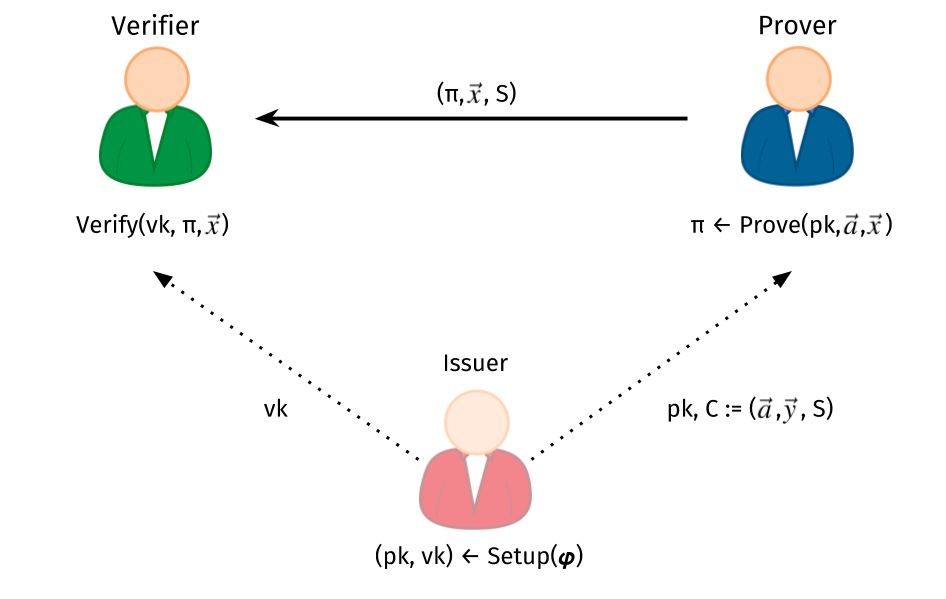}
  \caption{Actors and protocol primitives in a \thething{} use case.}
  \label{fig:overview}
\end{figure*}

Our contribution is \thething{}, a design and implementation for
non-interactive, privacy-preserving credentials through the use of
zero-knowledge proofs.
\thething{} allows users to act as provers and create proofs for any self-chosen statement over a credential issued to them without having to interact with another party.
This allows us to integrate \thething{} into a fully decentralized architecture
where non-interactive PP-ABCs are provisioned in a resilient, decentralized
delivery mechanism.
The core building blocks of our design are zero-knowledge succinct
non-interactive arguments of knowledge (zkSNARKs).
We present an evaluation of our implementation discussing the space and time
tradeoffs of such a system.
To illustrate the use of \thething{}, we show how it can be integrated into a
decentralized personal data sharing system.

\section{BACKGROUND}

zkSNARKs are a theoretical class of proofs which satisfy a specific set of
formal properties in order to realize a non-interactive zero-knowledge (NIZK)
proof~\cite{ben2013snarks}.
Its origin lies in the area of verifiable computation (VC) schemes.
There are two actors in VC: The prover and the validator.
Respective schemes allow to prove the correct evaluation of
a function given a set of inputs.
This is commonly advantageous in use cases where computation is offloaded onto
a third party because the actual computation is resource-intensive.
In this context, verifiable computation schemes are usually engineered to
provide an efficient way to verify computation results and put less emphasis on
the efficiency on the evaluation and proving processes.
In zkSNARKs, the idea is that such schemes can easily be extended to add a
zero-knowledge aspect to the computation:
The verifier of the computation must be able to verify the proof
(i.e. the result of the computation) without a private ``witness'' input.
The ``witness'' is only known to the prover, which is the entity that
performs the computation.
Popular verifiable computation schemes which allow to build zkSNARK proofs
include Pinocchio and a scheme by Groth et al.~\cite{parno2013pinocchio,groth2016size}.
The idea behind both approaches is that the verification of a result does not
require the function input.
Hence, the verifier is able to verify a computation result without knowledge
of what was actually the subject of the computation.

In the following, we define the functions and objects of a NIZK system,
consisting of a setup, credentials definition, circuit construction, proof
generation, verification, and a delivery mechanism.
We generalize the following high-level primitives of a zkSNARKs scheme:

\begin{align}
  Setup (\varphi) & \rightarrow (pk,vk)\label{eq:zksnarks1}\\
  Prove(pk, \vec{a}, \vec{x}) & \rightarrow \pi\label{eq:zksnarks2}\\
  Verify(vk,\pi, \vec{x}) & \rightarrow \{FALSE,TRUE\}\label{eq:zksnarks3}
\end{align}

Initially, we must establish a ``constraint system'' $\varphi$.
A constraint system is a set of linear constraints which are internally translated
into circuits by the zkSNARK scheme.
The constraint system is a blueprint that allows us to define ground truths and
derive the proving key $pk$ and verification key $vk$ using a $Setup()$ procedure.
The constraint system -- and consequently both $pk$ and $vk$ -- are public information and
are meant to be known by prover and verifier, respectively.
For a constraint system to be useful in our design, it must be constructed
in a way that supports proofs on credentials.
In order to achieve this, we define the setup and constraint system construction
in the design section of this paper.

In order to generate a proof $\pi$, the prover must supply the proof input
vectors $\vec{x}$ and $\vec{a}$ as well as the proving key $pk$.
While $\vec{x}$ is a public parameter, $\vec{a}$ is private and only known to
the prover.
To validate the proof $\pi$, a verifier uses the verification key $vk$ and the
public input vector $\vec{x}$ as inputs to the validation procedure.
The verification result is either valid (TRUE) or invalid (FALSE).

In the following, we use the above definitions of a zkSNARK scheme in order to
formalize the non-interactive credential system \thething{}.

\section{DESIGN}

\begin{figure*}[t]
  \centering
  \includegraphics[width=0.9\textwidth]{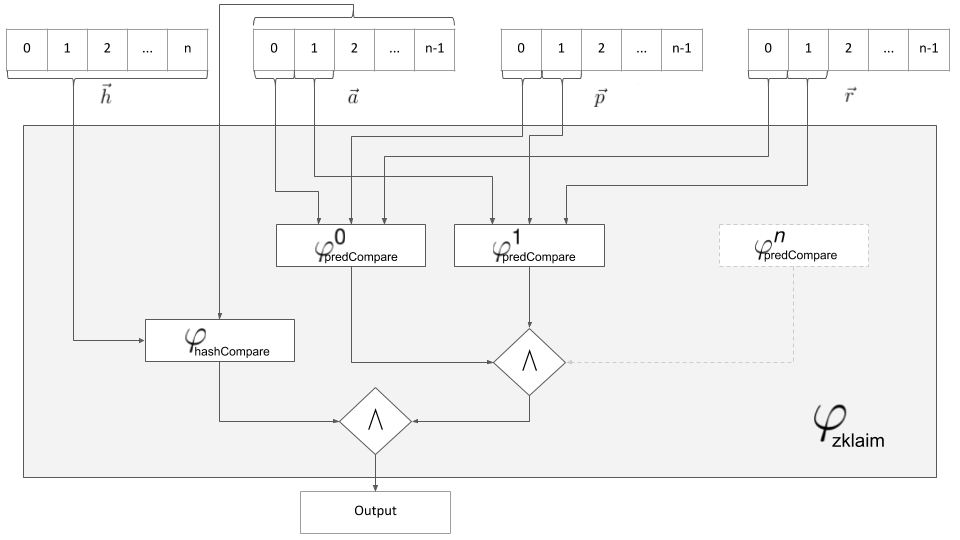}
  \caption{\thething{} constraint system $\varphi_{zklaim}$.}
  \label{fig:cs}
\end{figure*}

In the following, we present our design of \thething{} which satisfies the following
three requirements:

\textbf{Statements on credentials: }
\thething{} must allow users to generate proofs on third party issued
credentials.
The user must be allowed to freely choose the statement.
The verifier must be able to verify that the statement is true with respect
to the third party issued credential without the knowledge of the actual
credential value.

\textbf{Non-interactive presentation:}
\thething{} must allow verifiers to prove the correctness of a statement
non-interactively, e.g. without online interaction with the user or the credential
issuer.

\textbf{Selective disclosure:}
The user should have the option to selectively disclose credential values if
necessary.

First, in addition to the prover and verifier we define a third
actor: The credential issuer.
The issuer is a trusted third party that is issuing attribute-based credentials
(ABCs) to users which take the role of provers.
The credential contents are private and represent the private input vector
$\vec{a}$ of the proving procedure.
The prover is able to make any statement regarding its issued credentials and
create a proof $\pi$ which asserts that the statement is valid.
The verifier is an entity which requires the prover to prove the validity of a
certain statement.
In our use case, this statement is based on a specific attribute-based
credential.
The proof $\pi$ is zero-knowledge in that the verifier only learns whether or
not the statement on a credential is valid.
The contents of the credential are not disclosed to the verifier.
Figure \ref{fig:overview} illustrates our scenario including the generation of
the keys, a proof and its verification.
We use this illustration in the following sections to explain the design and
usage of \thething{}.

\subsection{Attributes and Credentials}

First, we define a \thething{} credential $\mathcal{C} := (\vec{a},\vec{y},S)$.
We define the input vectors $\vec{a}$ and $\vec{y}$ as bit vectors:
\begin{align}
  \vec{a} := & \vec{a_0} \mid \ldots \mid \vec{a_n}~where~\vec{a_i} \in \{0,1\}^*\\
  \vec{y} := & \vec{h_0} \mid \ldots \mid \vec{h_n}~where~~\vec{h_i} = hash(a_i)
\end{align}
$\vec{y}$ is the first and static part of the public input vector $\vec{x}$.
The other part of the input vector is variable and may be chosen by the user
as part of the proving process.
It is comprised of issuer-asserted user identity attributes such as date of
birth or email address.
$S$ is a signature over $\vec{y}$ and is created by the issuer.
The signature is created through traditional public-key cryptography.
This allows a verifier to establish trust from their set of trusted, third party
credential issuers to the credential $\mathcal{C}$ and verify its authenticity.

We note that $\vec{a}$ contains $n+1$ elements but there are only $n$
attributes while the last element is reserved.
This is a design choice for the following reason:
We define the last element $\vec{a_n}$ to be a unique identifier of the credential.
It is a random nonce generated by the credential issuer when the credential
$\vec{y}$ is issued.
The nonce ensures that the credential is unique across subjects even if
their attributes $\vec{a}$ are the same.

We expect an issuer to provide a mechanism that allows the user to retrieve
credentials $\mathcal{C}$ through a secure communication channel.
This transfer is out of scope of this work, but for web-based use cases it
can be realized through a traditional TLS channel in combination with
password-based user authentication.
Then, the transfer of the credential from the issuer to the user can be
performed through a simple download procedure.
The user then stores the credential in a wallet on a local storage under their
control.

\subsection{Constraint System and Keys}

As discussed in the background section, we must define a ``constraint system''
$\varphi$.
The entity responsible for creating the constraint system is the credential
issuer because it is the entity which is authoritative over what kinds of
attribute credentials it plans to issue to its users.
A \thething{} constraint system $\varphi_{zklaim}$ must be setup so that it
enables a prover to prove statements on credentials in the form $\mathcal{C}$.

Figure \ref{fig:cs} illustrates our circuit construction.

Constraint systems process input variables in an algebraic circuit and output
a boolean return value.
Hence, it is possible to combine multiple constraint systems into one new
constraint system.
In our design, we define the linear constraint system $\varphi_{zklaim}$ as a
combination of $n+1$ sub constraint systems:

\begin{align}
  \varphi_{zklaim} := \varphi_{hashCompare} \wedge (\bigwedge^n_{i = 0} \varphi^i_{predCompare})
\end{align}

The $hashCompare$ constraint allows the prover to verify that the user provided
private input vector $\vec{a}$ matches the credential $\mathcal{C}$ contents.
The second class of constraint systems are used to model, prove and verify
comparative statements on the private input $\vec{a}$.
For this the issuer must pre-determine the number $n$ of attributes that
$\vec{a}$ may contain as it determines the upper bound of sub constraint systems
of type $\varphi_{predCompare}$.
As illustrated in Figure \ref{fig:cs}, each $\varphi^i_{predCompare}$
constraint takes exactly one $a \in \vec{a}$ as input whereas the
$\varphi_{hashCompare}$ constraint system takes the whole input vector $\vec{a}$.
As constraint systems are rigid in this regard, a change in the number of
attributes requires a regeneration of the constraint system $\varphi_{zklaim}$.

\begin{figure*}[t]
  \begin{subfigure}[b]{0.45\textwidth}
  \includegraphics[width=\textwidth]{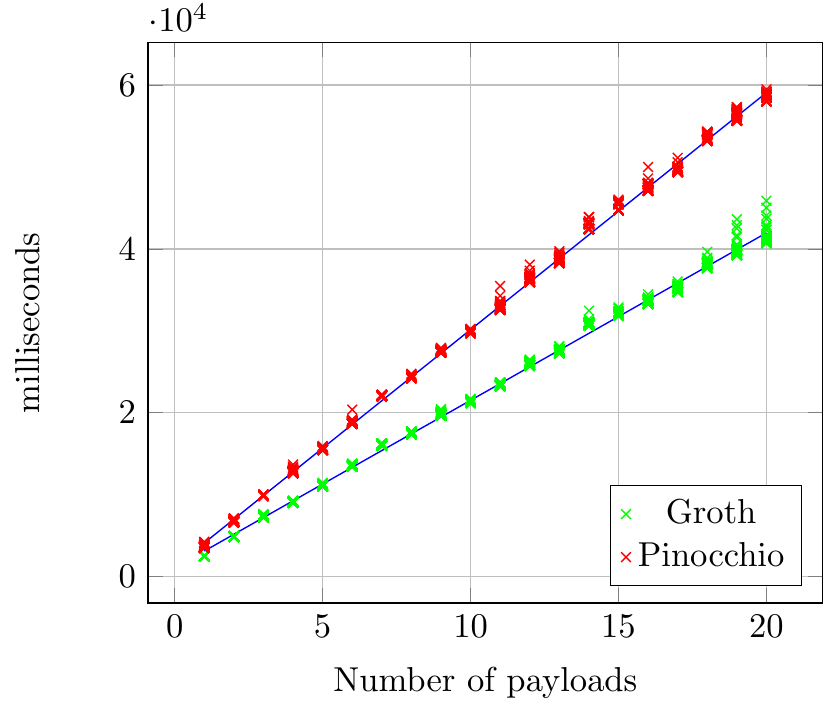}
  \end{subfigure}
  \qquad
  \begin{subfigure}[b]{0.45\textwidth}
  \includegraphics[width=\textwidth]{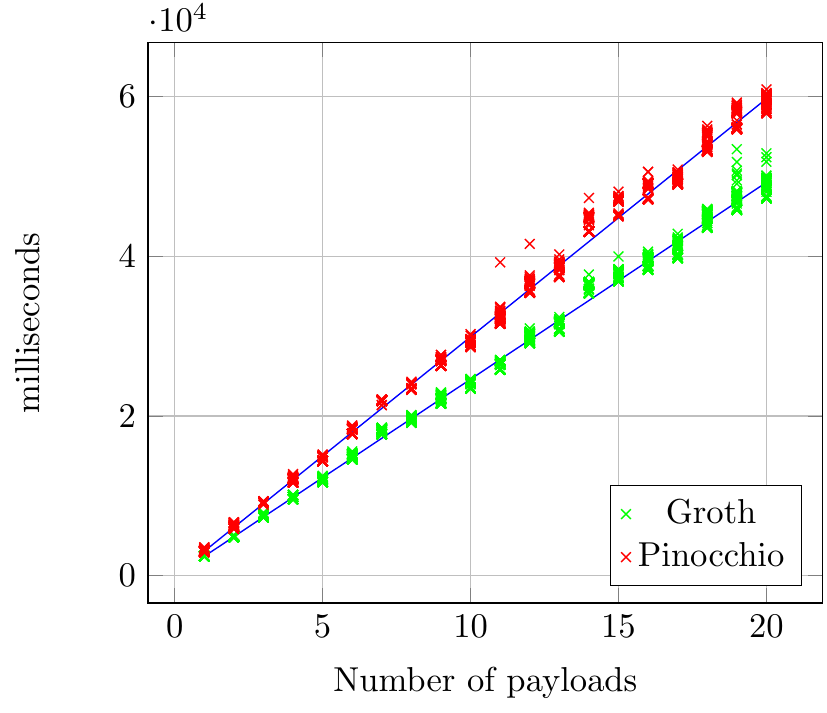}
  \end{subfigure}
  \caption{Left: time required to derive verification key $vk$ \& proving key $pk$ from the constraint system $\varphi_{zklaim}$. \\ Right: time required to create proof $\pi$ depending on the number of attribute payloads.}
  \label{fig:perf}
\end{figure*}

\subsection{Proving}

Using $\varphi_{zklaim}$ any entity is able to generate the public proving key
$pk$ and verification key $vk$ using the respective $Setup()$ procedure of the
zkSNARKs scheme.
The key $pk$ is used by the user in order to prove the validity of statements
on their attribute credentials.
Each $\varphi^i_{predCompare}$ may be used by the prover to
impose a predicate $\vec{p_i}$ with respect to a reference value $\vec{r_i}$.
The hashed attribute references in $\vec{y}$ are combined with the above into the
public proof input vector $\vec{x}$:
\begin{align}
  \vec{x} := \vec{y} \mid \vec {p} \mid \vec{r}
\end{align}

By default, each $\vec{p_i}$ is initialized as a no-op dummy operation which always
evaluates to true.
In order to create a statement on an attribute $\vec{a_i}$, the user sets the
predicate $\vec{p_i}$ to any combination of $<$, $=$ and $>$ or their respective
complements $\nless,\neq,\ngtr$.
This predicate is used in combination with a reference value $\vec{r_i}$ which
contains a value that the corresponding attribute $\vec{a_i}$ is to be checked
against with the predicate $\vec{p_i}$.
As an example, to create a proof input which is supposed to verify that a user
is born before a certain data, the reference value $\vec{r_i}$ for the
``data of birth'' attribute would be set to a certain timestamp in the past
which reflects the age barrier.
The position $n$ of the reference value is defined by issuer through
the constraint system.
The predicate is set to $\nless$.
Such a proof input $\vec{p}$ allows users to prove that they are over a certain age.

In general, $\vec{p}$ can be chosen arbitrarily by a prover.
However, $\varphi_{hashCompare}$ is used to import the requirement that
any prover must be able to provide a witness in the form of a pre-image
to $\vec{y}$, namely $\vec{a}$.
As already mentioned above, $\vec{a}$ -- and in particular
$a \in \vec{a}$ -- serves as a secret that the prover must present in
the proving process as part of a witness to the $\varphi_{hashCompare}$ constraint.
Hence, only the subject which is in possession of a credential $\mathcal{C}$
from the issuer is able to satisfy the constraint system.

In order to validate a proof $\pi$, the prover must apply the public proof input
$\vec{x}$, the proving key $pk$ as well
as the private input vector $\vec{a}$ to satisfy the constraint system
$\varphi_{zklaim}$ and generate a proof $\pi$.
The user generates a proof as follows:
\begin{equation}
  \pi \leftarrow Prove(pk, \vec{a}, \vec{x})
\end{equation}

The user is able to provide the hashes in $\vec{y}$ and the pre-image $\vec{a}$
from the credential $\mathcal{C}$ to satisfy the $\varphi_{hashCompare}$
constraint system.
The public input vector $\vec{x}$ is built using $\vec{y}$, the predicate
inputs vector $\vec{p}$ and the reference value vector $\vec{r}$ which
represent the statements made by the user on the attributes in $\vec{a}$.
We expect that the predicates are defined a priori, for example, through a
negotiation between the verifier and the user.
The verifier can request from the user to provide a specific proof including
certain predicates and thus defines the respective predicate variables $\vec{p}$
and $\vec{r}$.

After the user calculates the proof $\pi$, it can be presented to a verifier
along with $\vec{x}$.
We define a \thething{} context $(\pi, \vec{x}, S)$ which -- due to the
non-interactive nature of the proof -- can be persisted by the user and
non-interactively retrieved and verified by a verifier.

\subsection{Verification}

It is not necessary for a verifier to directly interact with the prover to
verify a proof $\pi$.
However, upon retrieving the \thething{} context $(\pi, \vec{x}, S)$ and before
the verification of $\pi$, the verifier must verify the signature $S$ over $\vec{y} \mid y_i \in \vec{x}$.
Using this information, the prover proceeds to retrieve the correct proving
key $pk$ from the trusted issuer and uses it to verify the proof $\pi$:

\begin{equation}
  result \in \{FALSE,TRUE\} \leftarrow Verify(vk, \pi, \vec{x})
\end{equation}

The verification function yields $TRUE$ if the user was able to provide inputs
to $\varphi_{zklaim}$ that satisfy the underlying constraint systems.
It is essential that verifiers check that the predicate inputs and reference
vectors $\vec{p}$ and $\vec{r}$, which are provided by the user as part of
the \thething{} context, are semantically what they expect them to be.
Especially if the verifier specifically requested a predicate to be proven,
such as ``$age \geq 18$'', the respective predicate (greater or equal) as well
as input variable to check against must be correctly set.

\section{IMPLEMENTATION AND EVALUATION}

Our reference implementation\footnote{\url{https://gitlab.com/kiliant/zklaim}, accessed 2019/01/08} is built on top of the \textit{libsnark}\footnote{\url{https://github.com/scipr-lab/libsnark}, accessed 2019/01/08} library.
zkSNARKs are based on verifiable computation schemes.
While libsnark supports a variety of different schemes including Pinocchio~\cite{parno2013pinocchio} we use the scheme of Groth~\cite{groth2016size} which is also readily available.
We settled on Groth because it exhibits better performance than the other schemes available in libsnark.

\begin{figure*}[t]
  \centering
  \includegraphics[width=0.6\textwidth]{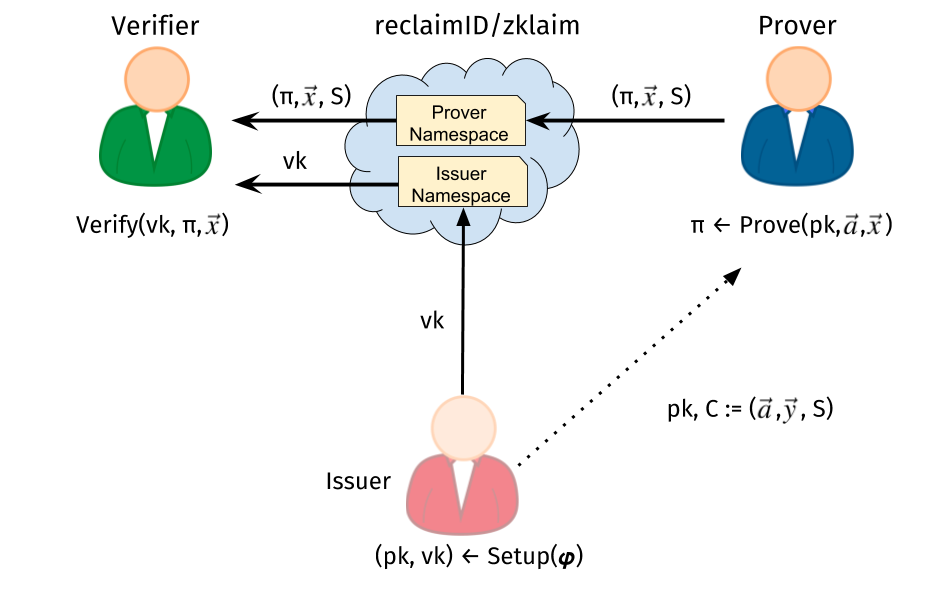}
  \caption{\thething{} integration with the reclaimID identity provider.}
  \label{fig:gnsarchitecture}
\end{figure*}

\subsection{Performance and Evaluation}

We evaluated the performance of \thething{} for issuing, proving and verification with respect to the number of attributes in a single credential.
Due to technical limitations imposed by the underlying constraint system we must use fixed size inputs to the constraint system.
We define a collection of attributes that fits into a single input as a
{\em payload}.
In our implementation, a single payload can hold up to five attributes.
Our test setup consisted of an Intel Core i7 7500U 3.2 GHz with 16 GB of RAM.
In Figure~\ref{fig:perf} we can see that the time it takes to construct the issuer constrain system increases linearly with the number of attribute payloads in our credential.
It takes roughly 2.5 seconds to build a constraint system that supports five attributes in a single payload and increases by the same amount for every additional set.
The time it takes a prover to construct a proof depending on the amount of payloads defined by the issuer constraint system can be found in Figure~\ref{fig:perf}.
We can see that just like the initial construction of the constraint system, proof construction time also increases linearly with the number of payloads.
Creating a proof in the case of a single payload takes roughly 2.4 seconds and increases by the same amount for every additional set.

While we evaluated the time it takes the verifier to validate proofs, the results
suggest that the impact is negligible:
In \thething{}, proof verification is simply a matter of evaluating a polynomial
function, measured times range below 10 milliseconds.
The issuer constraint system needs to be created only once at the beginning.
Proofs need to be constructed every time a verifier has a new request regarding the predicates it needs proven.
Once a proof for a combination of predicates exists, it can be stored and presented non-interactively to any concerned verifier.
In addition to the evaluation of performance, we also assessed the size of
proving and verification keys as well as proof size dependent on the number
of payloads.
The results can be found in Table~\ref{table:space}, where we find that the
minimum size of a proving key is roughly 8.65 MB.
With every payload -- i.e. every set of five attributes -- supported by the
issuer constraint system, the proving key increases in size by around 7 to 9 MB.
At 20 payloads, this results in a 174.51 MB proving key.
Due to this size constraint, issuers should either limit the number of
supported attributes or bootstrap dedicated constraint systems so that a prover
is only required to handle proving keys for attributes actually relevant to them.
At the same time, the verification key takes a minimum of 784 bytes and
increases by 150 to 200 bytes per payload.

In summary, the bulk of the space and time required in \thething{} must be
provided by the prover.
Proofs itself are of constant size at 137 bytes which is good news with respect
to the required storage footprint.
We can safely consider that most decentralized storage systems are capable of
accommodating \thething{} proofs.

\begin{table}[h]
  \centering
  \begin{small}
  \begin{tabular}{c | c | c | c}
    \hline
    \textbf{Payloads} & \textbf{$pk$ in MB} & \textbf{$vk$ in bytes} & \textbf{Proof in bytes} \\ \hline
    1           & 8.65  & 784  &  \\\cline{1-3}
    5           & 43.15  & 1543 &  \\\cline{1-3}
    10          & 86.94 & 2493 & 137 \\\cline{1-3}
    15          & 133.37 & 3443 &  \\\cline{1-3}
    20          & 174.51 & 4436 &  \\\hline
  \end{tabular}
  \end{small}
  \caption{Key and proof sizes depending on the number of payloads.}
  \label{table:space}
\end{table}

\subsection{Integration}

We designed \thething{} to specifically for decentralized identity provider services that require or support non-interactive presentation of identity attributes.
To publish and propagate \thething{} objects such as the issuer credential system, verification key and proofs, we propose the use of secure, decentralized identity provider systems based on name systems such as NameID~\cite{website:nameid} and reclaimID~\cite{schanzenbach2018reclaim}.

NameID~\cite{website:nameid}, is a blockchain-based identity system that allows
users to share identity attributes over the namecoin blockchain.
It features a standards-compliant delivery mechanism -- OpenID
Connect~\cite{website:oidc} -- but relies on a central rendezvous server.
The nature of both the OpenID Connect and blockchain architecture requires that
identity attributes can be presented without direct interaction between users
and relying parties.
This is partly due to the server-based architecture of OpenID Connect but also
a technical caveat of distributed ledgers.
In NameID, a central service in the form of an OpenID Connect server enforces
access control decisions made by users.

An alternative to NameID is reclaimID~\cite{schanzenbach2018reclaim}, which uses
the decentralized GNU Name System (GNS)~\cite{wachs2014censorship,wachs2014feasibility}.
reclaimID allows users to be completely sovereign over their own identities and selectively authorize access to identity attributes using attribute-based encryption (ABE).
This approach mitigates the issue of public records in the blockchain that we find in NameID.
reclaimID provides a fully decentralized storage and resolution mechanism for
identity attributes.
It enables relying parties, in our case represented by verifiers, to access
identity attributes without interacting with a trusted third party or the user.
Like NameID, reclaimID also features an OpenID Connect layer to allow standards-compliant integration into web services but does so without the use of a central server.
Instead, client-side software is used to emulate the OpenID Connect service on top the decentralized service infrastructure.

We have decided to integrate \thething{} into reclaimID due its more decentralized nature and some glaring shortcomings of NameID, such as public attribute records.
Given the strict size constraints of proving and verification keys, due to technical constraints of name systems we assume that verification keys must be exchanged
out-of-band.
However, since the authority over the issuer constraint system -- and with it the keys -- is the issuer itself and keys can be presumed to rarely change, out-of-band distribution using traditional means such as web servers is feasible.
On the other hand, distributing credentials and, more importantly, proofs using any of the above name system-based delivery systems is certainly possible.
Users create proofs and authorize verifiers to retrieve and verify them from the name system in an efficient, completely decentralized fashion.

In our implementation, the issuer publishes the \thething{} constraint system
$\varphi$, the verification key $vk$ and the proving key $pk$ in GNS.
This record is published in a namespace which is owned by the issuer.
This allows any prover to retrieve the issuer's constraint system and proving
key and to verify its integrity and use it as inputs in proving and verification
procedures.
Figure~\ref{fig:gnsarchitecture} illustrates the integration of \thething{}
with the reclaimID identity provider.
The prover shares the proving context including the proof $\pi$, the proof input
$\vec{x}$ and the credential signature $S$ with the verifier over reclaimID.
This is done by having the prover store the proving context as an attribute
record in reclaimID.
This attribute is shared with a verifier through an out-of-band authorization
protocol such as OpenID Connect.
Our reference implementation can be found online as part of the GNUnet
peer-to-peer framework\footnote{\url{https://gnunet.org/git/gnunet.git/tree/src/zklaim?h=zklaim}, accessed 2019/02/13}.

\section{RELATED WORK}

U-Prove is a digital credential technology that allows a prover to selectively
disclose claims issued by an issuer to a verifier~\cite{paquin2011u}.
The prover can choose which claims to present to the verifier and which to withhold.
Our approach differs from U-Prove in that it allows the prover to create a
claim using a predicate without interaction with the issuer.
For example, in U-Prove for provers to prove to an issuer that they
are ``over 18 years old'', they must request this statement as part of a
U-Prove token from the issuer.
In our design, the prover only requests the attribute -- e.g. ``is 24 years old''
-- as claim from the issuer.
The prover can use this attribute to create arbitrary proofs using predicates
based on this claim such as ``is not 20 years old'', ``is 24 years old'' or ``is over 18 years old''.

Identity Mixer (Idemix) is another sophisticated credential system that apart
from PP-ABCs also provides anonymity~\cite{camenisch2002design}.
It is already quite mature in that it already includes features such as
attribute predicates, revocation and selective disclose of attributes.
Further, Idemix allows a verifier to request disclosure of an attribute from the issuer.
What Idemix does not feature, is a non-interactivity property.
As such, ``offline'' presentation of a credential to a verifier is not possible
by design.
What Idemix gains from this restriction, is that a presented proof cannot be
re-used by the verifier to, e.g., impersonate the prover using the proof that
was presented to them.
This feature is only really relevant if the anonymity feature is also desired.
Currently, our system does not feature anonymity, so interactive sessions
between verifier and prover can be assumed to be authenticated.
The prover authentication can then be bound to the credentials in question,
for example through an attribute holding their public key.

The authors of UnlimitID, propose the use of
algebraic MACs for privacy-preserving credentials~\cite{isaakidis2016unlimitid}.
They propose a system which allows users to create pseudonyms in order to
make it impossible for the IdP to track users across relying parties.
UnlimitID supports the selective disclosure of user attributes.
However, it does not allow the user to prove the correctness of statements
on credentials without disclosing the credential value itself.
\section{SUMMARY AND FUTURE WORK}

In this paper we have presented \thething{}, a design for non-interactive
privacy-preserving credentials based on a non-interactive zero-knowledge protocol.
We have shown how zkSNARKs can be leveraged for decentralized identity provider
services.
We conducted performance evaluations of \thething{} to show that is can be used
in practice and where integrators must accommodate for additional resources.
Finally, we have integrated our \thething{} implementation into the
decentralized identity provider reclaimID.
This means improved privacy for reclaimID users if they choose to share
\thething{} proofs as attributes while at the same time providing relying parties
with strong assertions by trusted third parties.

As a next step, we plan to address shortcomings with current authorization
protocols such as OpenID Connect with respect to complex credentials such as
\thething{}.
OpenID Connect does not specify how relying parties can request special credential
types such as certificates, \thething{} or other third party asserted attributes.
This is due to the fact that the protocol was not originally designed to be
implemented on top of decentralized infrastructures.
However, in the wake of self-sovereign identity
systems~\cite{website:nameid,schanzenbach2018reclaim,website:sovrinWhitepaper},
this is a challenge in need of further research and development.

In future work we also plan to investigate how \thething{} can be used in the
Internet of Things.
Specifically, we plan on investigating how device can disclose metadata such as firmware versions to requesting parties in a minimal way.
This could allow services to query large fleets of devices for vulnerable
firmware versions without having all devices explicitly disclose the exact versions they run on.


\section*{ACKNOWLEDGMENTS}
This work was partially funded by the Fraunhofer Cluster Cognitive Internet Technologies.

\bibliographystyle{apalike}
{\small
\bibliography{references}}



\end{document}